\documentclass[]{emulateapj}

\newcommand{\s}{\mathrm{s}}
\newcommand{\Mpc}{\mathrm{Mpc}}
\newcommand{\km}{\mathrm{km}}
\newcommand{\asym}{\mathrm{asym}}
\newcommand{\vasym}{v_{\asym}}
\newcommand{\sasym}{\sigma_{\asym}}
\newcommand{\SFR}{\mathrm{SFR}}
\newcommand{\BzKgal}{\mathrm{BzK-}15504}
\newcommand{\pc}{\mathrm{pc}}
\newcommand{\kpc}{\mathrm{kpc}}
\newcommand{\yr}{\mathrm{yr}}
\newcommand{\Myr}{\mathrm{Myr}}
\newcommand{\Ha}{\mathrm{H}\alpha}
\newcommand{\vlos}{v_\mathrm{los}}
\newcommand{\Kasym}{K_\mathrm{asym}}
\newcommand{\sigmalos}{\sigma_\mathrm{los}}
\newcommand{\slos}{\sigmalos}

\newcommand{\gas}{\mathrm{gas}}
\newcommand{\LCDM}{\Lambda\mathrm{CDM}}
\newcommand{\kms}{\mathrm{\,km\,}\mathrm{s}^{-1}}

\newcommand{\Vvir}{V_{\mathrm{vir}}}

\newcommand{\Mstar}{M_{\star}}
\newcommand{\Mgas}{M_{\gas}}
\newcommand{\Msun}{M_{\sun}}

\shorttitle{Disk Galaxy Kinematics at $z\sim2$}
\shortauthors{Robertson and Bullock}

\begin{document}

\title{High-Redshift Galaxy Kinematics: Constraints on Models of Disk Formation}
\author{Brant E. Robertson\altaffilmark{1,2,4} and James S. Bullock\altaffilmark{3}}

\altaffiltext{1}{Kavli Institute for Cosmological Physics, and Department of Astronomy and
Astrophysics, University of Chicago, 933 East 56th Street, Chicago, IL 60637, USA}
\altaffiltext{2}{Enrico Fermi Institute, 5640 South Ellis Avenue, Chicago, IL 60637, USA}
\altaffiltext{3}{Center for Cosmology, Department of Physics and Astronomy, University of California, Irvine, CA 92697}
\altaffiltext{4}{Spitzer Fellow}

\begin{abstract}
Integral field spectroscopy of galaxies at
redshift $z\sim2$ has revealed a population of
early-forming, rotationally-supported disks. These high-redshift systems provide a potentially important
clue to the formation processes that build disk galaxies in the universe.
A particularly well-studied example is
the $z=2.38$ galaxy $\BzKgal$,  which was
shown  by  \cite{genzel2006a} to be a rotationally supported disk despite the fact that its high
star formation rate and short gas consumption timescale require a very rapid acquisition of mass.
Previous kinematical analyses have suggested that
$z\sim2$ disk galaxies like $\BzKgal$ 
did not form through mergers because their
line-of-sight velocity fields display low levels of asymmetry.
We perform the same kinematical analysis on a set of simulated disk galaxies formed in 
gas-rich mergers of the type that may be common at high redshift, and show
 that the remnant disks display low velocity field asymmetry and
satisfy the criteria that have been used to classify high-redshift galaxies as disks observationally.  
Further, we compare one of our remnants to the bulk properties of $\BzKgal$ and
show that it has a star formation rate, gas surface density, and
a circular velocity-to-velocity dispersion ratio that matches $\BzKgal$ remarkably well. 
We suggest that observations of high-redshift
disk galaxies like $\BzKgal$ are consistent with the hypothesis that gas-rich mergers
play an important role in disk formation at high redshift.

\end{abstract}


\section{Introduction}
\label{section:introduction}

Understanding the formation of disk galaxies remains a primary but elusive
goal in cosmology.  The standard scenario for disk formation
entails the quiescent, dissipational
collapse of rotating gas clouds within virialized dark matter halos
\citep{white1978a,fall1980a,blumenthal1984a}, and this picture
 successfully explains the angular momentum
content, size, and kinematical structure of observed systems 
\citep[e.g.,][]{mo1998a}.  However, 
CDM-based
cosmological simulations of disk galaxy formation have had difficulty
producing high angular momentum disks without large bulges \citep[e.g.,][]{navarro1994a,navarro2000a}, 
almost certainly in part due to the fact that 
mergers are common in Cold Dark Matter (CDM) cosmologies
\citep[][and references therein]{stewart08a}.  
More recent simulations have faired better by
including  strong feedback, which suppresses star formation and thereby
reduces the collisionless heating associated with stellar mergers
\citep[e.g.,][]{abadi2003a,sommer-larsen2003a,robertson2004a,governato2004a}.

Motivated by these results,
\citet[][hereafter, R06a]{robertson2006a} presented a supplemental ``merger-driven''
scenario for the cosmological disk galaxy formation where extremely gas-rich mergers at high-redshift lead to rapidly-rotating remnant systems with large gaseous and stellar disks.
Building on work by \cite{barnes2002a}, \cite{brook2004a}, and \cite{springel2005a},
R06a used a suite of
hydrodynamical simulations to study gas rich mergers of various gas fractions, mass ratios, and
orbital properties, and  
concluded that hierarchical structure formation and the regulation of
star formation may work in concert to build disk galaxies through gas-rich mergers.
Subsequent cosmological simulations  have reported the
formation of Milky Way-like disk galaxies in gas-rich mergers \citep{governato2007a},
and \cite{hopkins2008a} have examined merger remnants in order to gain a
phenomenological understanding of disk survival in gas-rich mergers.
Given that high gas fractions are the essential ingredient in this scenario, observations 
at high-redshift (when gas fractions were higher and mergers more common) provide an important testing ground
for models of disk galaxy formation.

Only recently have detailed observations of high-redshift disks
been possible.
 \cite{erb2003a} measured the rotation curves of UV-selected
$z\sim2$
galaxies \citep[the BM/BX sample,][]{adelberger2004a,steidel2004a} with $\Ha$ slit spectroscopy \citep[see also][]{erb2006a}.
\cite{forster-schreiber2006a} used the SINFONI integral field spectrograph \citep{eisenhauer2003a}
to measure spatially-resolved kinematics of 
14 BM/BX galaxies, and found rotating systems with large specific angular momenta
but significant random motions ($v/\sigma\sim2-4$).

Among the best-studied high-redhift disks is  $\BzKgal$ at $z=2.38$, which was 
found using the $BzK$ photometric selection technique \citep{daddi2004a}.
 \citet[][hereafter G06]{genzel2006a}
used the SINFONI spectrograph with  adaptive optics on the
Very Large Telescope (VLT) to measure the kinematic structure
$\BzKgal$ with an incredible spatial resolution 
of 0.15" ($\sim1.2\,\kpc$ at $z \simeq 2.4$).  
These observations revealed
massive ($\Mstar \sim 8\times10^{10}\Msun$, $\Mgas \sim 4\times10^{10}\Msun$) rotating disk galaxy with a large
star formation rate ($\SFR\sim140\Msun\yr^{-1}$) and substantial random motions ($v/\sigma\sim3$).
The inferred gas consumption timescale for $\BzKgal$ is only $\sim285~\Myr$,
or about one-tenth the age of the universe at $z\sim2.38$,  and implies a remarkably 
rapid acquisition of its mass.  Nonetheless G06 conclude that $\BzKgal$ is
not a merger remnant because it displays 
velocity fields similar to those expected for
quiescent disks.

The existence of high-redshift disks that have not experienced large mergers may be
surprising in light of $\Lambda$CDM expectations.
 The merger rate per dark matter halo is predicted to be 
a factor of $\sim15$ times higher at $z \sim 2.4$ than at $z\sim0$ \citep{fakhouri2008a}.
Approximately $10 \%$ of galaxy-size halos at $z \simeq 2.4$
should have undergone a nearly one-to-one merger ( $>1:1.25$)  in the last $\sim500~\Myr$,
and  $\sim 50 \%$ should have experienced a $>1:3$ merger in the last $\sim 1$ Gyr (Stewart, Bullock et al., in preparation).
However, these mergers are expected to be gas rich.   Observations of galaxies at $z\sim2$ imply gas fractions
$\sim10-20$ times higher than for nearby galaxies at fixed stellar mass 
\citep{erb2006a,gavazzi2008a}.   These two pieces of information motivate us to
consider gas-rich mergers as a means to explain $z\sim 2$ galaxies with young stellar populations, 
short gas exhaustion timescales, and disk-like kinematics.

 Any model that attempts to explain the properties of high-redshift
disks must not only reproduce their bulk properties
 (stellar masses, rotational velocities, gas fractions, star formation rates, etc.)
 but also their detailed kinematic properties.
An important kinematic metric for classifying high-redshift galaxies as disks was developed by
\citet[][hereafter S08]{shapiro2008a}, who extended the ``kinemetry'' technique of \cite{krajnovic2006a}
in order to examine the velocity fields 
of $z\sim2$ galaxies from
 \cite{forster-schreiber2006a} and G06.  
Specifically, S08 provided a straightforward means to classify $z\sim2$ galaxies as disks
 based on the symmetry of their velocity fields.

The purpose of this {\em Letter} is to determine whether galaxies formed
in a gas-rich mergers can match the observed properties of $z \sim 2$ disks.
We describe our
simulations in \S \ref{section:methodology} and 
the kinematical analysis in \S \ref{section:kinemetry}. 
 In \S \ref{section:results} we demonstrate
 that a set of gas-rich merger remnants satisfy the
 disk galaxy kinematical classification developed by S08, and
 compare one of the remnants in particular to the detailed
 properties of  $\BzKgal$.
Overall, we find good agreement between the properties
of simulated disk galaxies formed in gas-rich mergers and 
observed $z\sim2$ disks like $\BzKgal$.
We assume a $\LCDM$ cosmology with $\Omega_{m}\simeq0.3$, $\Omega_{\Lambda}\simeq0.7$, and a Hubble parameter of $H_{0} = 70 \km\s^{-1}\Mpc^{-1}$.

\section{Simulation Methodology}
\label{section:methodology}

The simulations studied here are culled from the simulation suite
presented in R06a.
For details of the simulation methodology we refer the reader to
\cite{springel2003a}, \cite{springel2005b}, \cite{springel2005c}, and \cite{robertson2006b,robertson2006c},
but a summary follows.  The simulations are
calculated using the N-body and Smoothed Particle Hydrodynamics code GADGET2 
\citep{springel2001a,springel2005c}.  Each binary merger occurs during a parabolic 
encounter between otherwise isolated disk galaxies.
The progenitor galaxy models contain gaseous and stellar disks embedded in massive dark matter halos,
with structural parameters scaled appropriately for disk galaxies in the $\LCDM$ cosmology 
\citep[e.g.,][]{mo1998a,springel2005b}.
The gas and stars are initialized
in exponential disks with a scale height $H_{d}$ to scale length $R_{d}$ ratio of $H_{d}/R_{d}=0.2$,
while the dark matter halo is initialized as rotating ($\lambda=0.033$) \cite{hernquist1990a} distribution with an
approximate \cite{navarro1996a} concentration of $c_{\mathrm{NFW}}=R_{200}/R_{s}=9$
(the simulations of R06a were initialized for concentrations typical of $z\sim0$).
We note that while
the concentration is higher than that expected for
similar mass halos at $z\sim2$ \citep[e.g.,][]{bullock2001a}, the integrated mass within the
central $\sim10~\kpc$ is approximately correct.
The evolution in halo concentrations with redshift is 
primarily driven by growth in halo virial radii with time, while
halo central densities remain approximately constant
\citep[see, e.g., Figure 18 of][]{wechsler2002a}.
The progenitor galaxy models each contain 40,000 gas,
40,000 stellar disk, and 180,000 dark matter halo particles,
with a gravitational softening of $\epsilon = 100h^{-1}\pc$.

The prescription for star formation
and interstellar medium (ISM) physics in the simulations follows the \cite{springel2003a} model 
that implements a sub-resolution treatment of supernova feedback and 
the multiphase ISM.  
Star formation operates in the dense ISM, following a \cite{schmidt1959a} law
where the star formation timescale decreases with the local dynamical time.
The simulations of R06a utilize the strong supernova feedback supplied by the
\cite{springel2003a} multiphase ISM model
to suppress star formation sufficiently during a merger to allow remnant disks to form.
The simulations do not include an ultraviolet (UV) background, but we
note that the increased UV background at $z\sim2$ would tend to suppress
the overall gas consumption during the merger and may
act to increase the rotational support of the final remnants.
For comparison with the observations of $\BzKgal$ by G06,
we focus on a single
equal-mass coplanar binary merger of 
disk galaxies with virial velocities $\Vvir=160\kms$ and a pericentric passage distance of
$r_{\mathrm{peri}} \approx 2R_{d} = 8.6 \kpc$. 
The model is simulation "GE" in the study by 
R06a (see their Table 1)
and
is similar to the simulation studied by \cite{springel2005a}.
At the time of final coalescence $\Delta t\approx 450~\Myr$ after the first passage, the 
merging system is $\approx60\%$ gas and experiences a large burst of star formation.
The system averages $\SFR\approx150\Msun\yr^{-1}$ until reaching a quiescent rate of
$\SFR\approx15\Msun\yr^{-1}$ approximately $285~\Myr$ later.
We analyze
the simulation on the declining peak of the vigorous star formation burst, 
approximately $140~\Myr$ after the start of the final coalescence (roughly $100~\Myr$ before the
third panel of Figure 1 in R06a).
By this stage, a large gaseous
disk has formed from the residual angular momentum of the merger.  Eventually, the system forms a
rotationally-supported stellar component with a majority of its stellar mass distributed in
thin and thick disks (see Figures 2 and 3 and Table 2 of R06a).

\begin{figure*}
\figurenum{1}
\epsscale{0.7}
\plotone{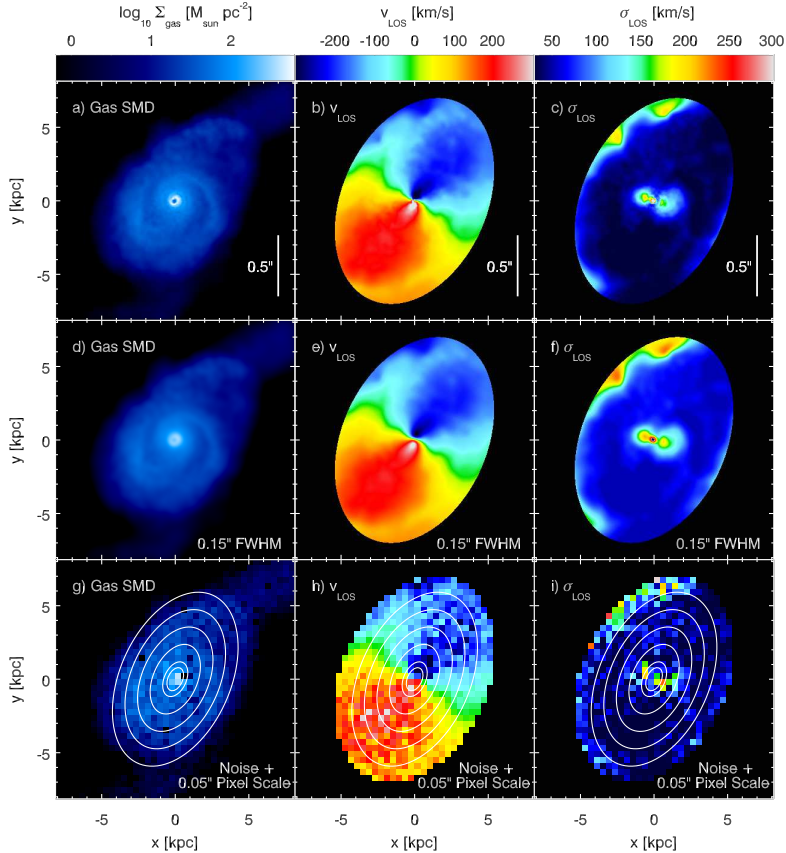}
\caption{\label{fig:vel_field}
Density and velocity fields of a simulated disk galaxy formed in a gas-rich merger.
The disk is rotated to an inclination of $i=48\deg$ and a position angle $\mathrm{PA}=24\deg$ West of North
for comparison with  the redshift $z=2.348$ galaxy $\BzKgal$ (G06, Figure 3).
The scale bar in the upper panels shows the relative size of 0.5" at $z=2.348$.
The left, middle, and right columns show the
surface mass density (SMD), velocity ($\vlos$), and velocity dispersion ($\sigmalos$), respectively,
 for gas associated $\Ha$ emission.  
 The color scales display the range of SMD, velocity, and velocity dispersion for all rows. The upper row (a-c)
provides  high-resolution maps of the simulation, while the middle row (d-f) is smoothed using a 0.15" FWHM gaussian, typical of VLT resolution with adaptive optics. The bottom row (g-i) shows the smoothed fields binned to a 0.05" pixel scale,  chosen to match the capabilities of the SINFONI integral field spectrograph \citep{eisenhauer2003a}. 
Gaussian noise is added to the binned maps to match the typical velocity errors reported by S08.
The kinemetric analysis designed by \cite{krajnovic2006a} and S08 is performed on the  pixelated $\vlos$ (panel h) and $\sigmalos$ (panel i) maps and
yields an asymmetry parameter of   $\Kasym=0.1$. S08 classify galaxies with $\Kasym<0.5$ as disks.
}
\end{figure*}

\section{Kinematical Analysis}
\label{section:kinemetry}

The kinematical properties of the simulated merger remnants are analyzed using a
method designed to closely approximate the integral field spectroscopy studies of 
\cite{forster-schreiber2006a}
and G06.
Figure \ref{fig:vel_field} illustrates this analysis for the GE remnant from R06a discussed in \S 2. 
The disk is inclined to $i=48\deg$ and
rotated to a major kinematic axis position angle of $\mathrm{PA}=24\deg$ West of North in order to
match $\BzKgal$ (see G06, their Figure 3).
Star-forming gas and diffuse gas with a temperature in 
the range $3\times10^{3}\leq T\leq3\times10^{4}$K are selected to approximate the observed $\Ha$-emitting
gas and the size of the simulated galaxy is scaled to the redshift of
$\BzKgal$, $z=2.38$ (see the angular scale bar in panels a-c of Figure \ref{fig:vel_field}).
 The three columns of Figure \ref{fig:vel_field}
show gas surface mass density (SMD),  line-of-sight velocity $\vlos$, and
velocity dispersion $\slos$ maps of the simulated disk gas displayed at three
different smoothing scales.  The upper row (a, b, c) is rendered at high resolution.
Note the clear disk morphology (a) and disk-like velocity field (b), while
the random motions (c) are larger than for thin disks in the local
universe (we measure, $\vlos/\sigmalos \sim 2-4$ in the remnant gas disk).
In the middle row (d, e, f) the fields are  smoothed by a gaussian kernel with a FWHM$=0.15"$ 
in order mimic the angular resolution obtained by G06.~\footnote{Note that most  kinematical studies of $z\sim2$ galaxies
have $\sim0.5"$ resolution (see  \citealt{forster-schreiber2006a} and S08). } 
In the lower row (g, h, i)
the smoothed maps are then binned to the $0.05"$ pixel scale of the SINFONI observations by 
G06 and gaussian noise is added to match the typical velocity errors reported by S08 (see their Figure 6).

\begin{deluxetable*}{l|cc} 
\tablecolumns{3} 
\tablewidth{0pc} 
\tablecaption{Properties of $\BzKgal$ vs. Simulated Disk Merger Remnant} 
\tablehead{
\colhead{Property} & \colhead{$\BzKgal^{1}$} & \colhead{Disk Remnant$^{2}$}
\label{table:properties}}
\startdata 
Dynamical Mass (r$<1.1$") & $1.1\pm1\times10^{11}\Msun$ & $1.2\times10^{11}\Msun$ \\
Stellar Mass            & $7.7(+3.9,-1.3)\times10^{10}\Msun$ & $7.3\times10^{10}\Msun$ \\
Gas Mass                & $4.3\times10^{10}\Msun$ & $3.8\times10^{10}\Msun$ \\
Average $\Sigma_{\gas}$ & $350\Msun \pc^{-2}$ & $306\Msun \pc^{-2}$ \\
Average $\Sigma_{\SFR}$ & $1.2\Msun \yr^{-1} \kpc^{-2}$ & $1.1 \Msun \yr^{-1} \kpc^{-2}$ \\
$v_{c}/\sigma$              & $3\pm1$                       & $3.2\pm2$\\
Gas Disk Scale Length           & $4.5\pm1\,\kpc$                 & $2.6\,\kpc$\\
\enddata 
\tablerefs{
(1) \cite{genzel2006a}; (2) \cite{robertson2006a}
}
\end{deluxetable*}

The kinemetry analysis developed by 
\cite{krajnovic2006a} is then applied to the pixelated $\vlos$ and $\slos$ maps.  
This
method calculates an expansion of the velocity field along an ellipse with semi-major axis $a$
as a function of azimuthal angle $\psi$ of the form
\begin{equation}
\label{equation:kinemetry}
K(a,\psi) = A_{0}(a) + \sum_{n=1}^{n_{\mathrm{max}}} \left[ A_{n}(a) \sin n\psi +  B_{n}(a) \cos n\psi \right],
\end{equation}
\noindent
where $n_{\mathrm{max}}$ is a maximum term in the expansion and the 
coefficient $B_{1}$ describes the circular velocity of the system.
For high-quality kinematical data 
of nearby galaxies \citep[e.g., from SAURON, see][]{bacon2001a,cappellari2006a}, the kinemetry analysis provides a detailed exposition of the kinematical structure of 
galaxies \citep{krajnovic2006a}.  The comparatively coarser data available 
for $z\sim2$ galaxies motivated S08 to use
 a mean kinemetric analysis where the asymmetries are characterized by
averages of the amplitude coefficients in Equation \ref{equation:kinemetry}.  Re-writing
the kinemetry coefficients as
$k_{n} = \sqrt{A_{n}^{2} + B_{n}^{2}}$,
S08 introduced average velocity and velocity dispersion
asymmetry parameters
\begin{equation}
\label{equation:vasym}
\vasym  = \left< \frac{\sum_{n=2}^{5} k_{n,v}/4}{B_{1,v}}\right>_{r},  \hspace{.1in} \sasym  = \left< \frac{\sum_{n=1}^{5} k_{n,\sigma}/5}{B_{1,v}}\right>_{r},
\end{equation}
\noindent
where the subscripts $v$ and $\sigma$ refer to quantities calculated
from the $\vlos$ and $\slos$ maps, and the subscript $r$
indicates that the average is performed over the ellipses at different
radii used in the kinemetry analysis.  We use the kinemetry code
made available by D. Krajnovic
to calculate $\vasym$ and $\sasym$ for the simulated disk remnant. 
The kinemetry 
is performed over six ellipses 
separated by $\sim3$ pixels in the map, roughly corresponding to the 
resolution of the simulated image 
to minimize beam smearing effects  
\citep[e.g.,][]{van_den_bosch2000a}.  
The ellipse position angles and aspect ratios are
fixed to reflect the best fit values for $\BzKgal$ adopted from G06.
The number, 
spacing, and orientation of the ellipses 
can change the precise values of $\vasym$ and $\sasym$, but do not alter 
the conclusions or interpretations of this work.
Increasing the level of noise in the simulated maps will tend to
increase the values of $\vasym$ and $\sasym$, and would likely
degrade the capability of the simulations to match the observations
rather than improve our results.
We note that while the kinemetry method \cite{krajnovic2006a} has previously been used analyze simulated merger
remnants
\citep{jesseit2007a,kronberger2007a},  
only very gas-rich mergers are expected to be associated
with disk formation and kinemetry has not yet been performed on such systems.

\section{Results and Discussion}
\label{section:results}

The bottom row of Figure \ref{fig:vel_field}
shows the SMD, $\vlos$, and $\slos$ maps used
to perform the kinematical analysis of the remnant
disk.  The average kinemetric parameters $\vasym$
and $\sasym$ (Equation 2)
are measured from the velocity maps.  For the disk merger remnant
plotted in 
Figure \ref{fig:vel_field} we find 
$\vasym=0.076$ and $\sasym=0.063$, which reflect the  
relative symmetry the velocity fields.
The combined asymmetry parameter
$\Kasym\equiv\sqrt{\vasym^{2} + \sasym^{2}}$
provides a global characterization of asymmetry 
in the kinematical structure of the galaxy, and
S08 suggest $\Kasym<0.5$ as an
observational criterion for distinguishing between
disk galaxies and mergers.  The disk remnant examined
in Figure \ref{fig:vel_field} satisfies this 
criterion with $\Kasym=0.1$, even though it was
formed in an equal mass, gas-rich merger.  
We have performed the same analysis on a range of 
other simulated merger remnant disk galaxies from R06a.  
In each case, we examine the merger product soon after 
the merger and $150~\Myr$ after the time of the final coalescence.
We find similar results for disk remnants formed in gas-rich 
major polar mergers (R06a simulation GCoF, $\Kasym=0.14$),
major inclined mergers (R06a simulation GCoO, $\Kasym=0.15$),
and minor (1:8 mass ratio) coplanar mergers (R06a simulation FCm, $\Kasym=0.16$).
If real galaxy analogues of these simulated galaxies were observed with SINFONI, 
a kinematical analysis would likely classify them (correctly) as disks.  
We emphasize that in each case the gas disk develops rapidly, with $v/\sigma$ 
and $\Kasym$ both declining by factors of $\sim2-3$ within $\sim100~\Myr$ of 
the height of merger coalescence.

While both the simulated disk merger remnants and observed $z\sim2$ disk
galaxies satisfy the observational criterion of S08 designed
to identify disk galaxies, the simulations should also closely match other 
observed properties
of $z\sim2$ disks.  Consider the galaxy $\BzKgal$ observed by  G06,
which has an inferred star formation rate of
$\SFR\approx140^{+110}_{-80}\Msun \yr^{-1}$ and a dynamical mass of 
$M_{\mathrm{dyn}}=1.1 \pm 0.1 \times10^{11}\Msun$
within a radius of $r\lesssim 9\kpc$.  The simulated disk remnant analyzed
in Figure \ref{fig:vel_field} has an average star formation rate of 
$\SFR\approx150 \Msun \yr^{=1}$ over the previous $100~\Myr$ and a dynamical
mass of $M_{\mathrm{dyn}}=1.2\times10^{11}\Msun$.  The inferred 
stellar and gas masses for $\BzKgal$ are 
$\Mstar = 7.7^{+3.9}_{-1.3}\times10^{10}\Msun$
and $\Mgas = 4.3\times10^{10}\Msun$, while the disk remnant has
a stellar mass of $\Mstar = 7.3\times10^{10}\Msun$ and a gas mass of
$\Mgas = 3.8\times10^{10}\Msun$.  Table \ref{table:properties} compares
the salient properties of $\BzKgal$ and the simulated disk remnant, and
they are similar in each case. 
Overall, the galaxy $\BzKgal$ displays properties that
are remarkably similar to the example simulated disk galaxy 
merger remnant shown in Figure \ref{fig:vel_field}.

We note that the properties of $z\sim2$ disks like $\BzKgal$ are
rather extreme compared with local, isolated disk galaxies.  
The gas surface density,
star formation rate, and random motions ($v/\sigma$) 
of $\BzKgal$ are roughly an order of
magnitude larger than for the Milky Way, even as their 
rotational velocities are similar and their stellar masses differ
by less than 50\% \citep[e.g.,][]{blitz1980a,rana1991a,kent1991a,klypin2002a,levine2006a}.
Our simulated disk remnant displays similar properties (see Table \ref{table:properties}) 
owing to the dynamical effects of the merger from which it formed. 
It is possible that such a pre-formed (hot) disk would tend to settle into a more
classical Milky-Way type thin disk at late times if the pre-formed stars adjust to 
to the infall of (quiescent) gas-disk material, as might be expected in a more classical
stage of disk formation. Subsequent gaseous infall from the intergalactic medium
may also help prolong star formation in
the remnant and increase the gas disk scale length.

While the observed properties of $z\sim2$ disks resemble simulated disk systems
formed in gas-rich mergers, other viable explanations for their origin exist.
Notably, \cite{bournaud2007a,bournaud2008a} have suggested that high-redshift
disk galaxies undergo a ``clump-cluster'' phase where large-scale gravitational
instability leads to a turbulent, clumped gaseous distribution \citep[see also][]{noguchi1998a,noguchi1999a}.  
An analysis of clump-cluster galaxy models using kinemetry method of \cite{krajnovic2006a} and S08
would provide a useful test of this scenario.

As the quantity and quality of kinematical data for $z\sim2$ disk galaxies improve,
the observational constraints will better distinguish between various 
models for their formation.  These observations will be essential for determining
how disk galaxies are assembled over the history of cosmic structure formation.

\acknowledgements

BER gratefully acknowledges support from a Spitzer Fellowship through
a NASA grant administrated by the Spitzer Science Center, and
the Kavli Institute for
Cosmological Physics at the University of Chicago.  JSB was supported by
NSF  grants and the Center for Cosmology at UC Irvine.
We also thank D. Krajnovic for making his kinemetry analysis code
easily accessible and A. V. Kravtsov for useful comments and suggestions.
We appreciate helpful comments and suggestions from the anonymous referee.

\clearpage

\clearpage

\end{document}